\documentclass[%
 reprint,
 amsmath,amssymb,
 aps,
]{revtex4-2}

\usepackage{graphicx}
\usepackage{dcolumn}
\usepackage{bm}
\usepackage{ulem}

\usepackage{color}
\usepackage{xcolor}
\usepackage{booktabs}
\usepackage{amsmath}
\usepackage{commath}

\newcommand{\grad}{\boldsymbol{\nabla}}

\newcommand{\crsc}[1]{\left[ #1 \right]}

\newcommand{\rnl}{\mathrm{Re}}

\newcommand{\cm}[1]{\textcolor{black}{{#1}}}

\begin{document}

\preprint{APS/123-QED}

\title{Nematic order condensation and topological defects in inertial active nematics}

\author{Roozbeh Saghatchi}
 \affiliation{Faculty of Engineering and Natural Sciences, Sabanci University, Tuzla, 34956 Istanbul, Turkey. \\
Integrated Manufacturing Technology Research \& Application Center, Sabanci University, Tuzla, 34956 Istanbul, Turkey. \\
Composite Technologies Center of Excellence, Sabanci University-Kordsa, Pendik, 34906 Istanbul, Turkey. \\
}%
\author{Mehmet Yildiz}%
\email{E-mail: mehmet.yildiz@sabanciuniv.edu}
\affiliation{%
Faculty of Engineering and Natural Sciences, Sabanci University, Tuzla, 34956 Istanbul, Turkey. \\
Integrated Manufacturing Technology Research \& Application Center, Sabanci University, Tuzla, 34956 Istanbul, Turkey. \\
Composite Technologies Center of Excellence, Sabanci University-Kordsa, Pendik, 34906 Istanbul, Turkey. \\
}%


\author{Amin Doostmohammadi}
\email{E-mail: doostmohammadi@nbi.ku.dk}
\affiliation{
 Niels Bohr Institute, University of Copenhagen, Copenhagen, Denmark.
}%



\begin{abstract}
Living materials at different length scales manifest active nematic features such as orientational order, nematic topological defects, and active nematic turbulence. Using numerical simulations we investigate the impact of fluid inertia on the collective pattern formation in active nematics. We show that an incremental increase in inertial effects due to reduced viscosity results in gradual melting of nematic order with an increase in topological defect density before a discontinuous transition to a vortex-condensate state. The emergent vortex-condensate state at low enough viscosities coincides with nematic order condensation within the giant vortices and the drop in the density of topological defects. We further show flow field around topological defects is substantially affected by inertial effects. Moreover, we demonstrate the strong dependence of the kinetic energy spectrum on the inertial effects, recover the Kolmogorov scaling within the vortex-condensate phase, but find no evidence of universal scaling at higher viscosities. The findings reveal new complexities in active nematic turbulence and emphasize the important cross-talk between active and inertial effects in setting flow and orientational organization of active particles.
\end{abstract}

\maketitle


\section{\label{sec:int}Introduction}
Active matter describes systems consisting of elements that are capable of converting energy - from internal mechanisms or extracted from the surrounding - into a mechanical work. As such, active matter inherently operates far from thermodynamic equilibrium and is an integral element for many living systems on the scales ranging from animal herds to suspensions of bacteria and cellular assemblies~\cite{Marchetti13,Prost15,Needleman17,Doost18}.

The continuous injection of energy at the level of individual active particles results in many interesting physical characteristics, including the emergence of collective self-organization and coherent chaotic flows termed active turbulence~\cite{wensink2012meso,Frey15,doost2017,doostmohammadi2017onset,Urzay17,alert2021active}. Unlike classical turbulence that is driven by external forcing, active turbulence initiates from the inherent hydrodynamic instability of active matter that is driven at the scale of individual elements~\cite{simha2002hydrodynamic,martinez2019selection}. As a result of this hydrodynamic instability the active matter exhibits flow patterns with a characteristic length that is larger than the size of constituent particles~\cite{Marchetti13,klotsa2019above}, and flow vortices and jets with a significantly larger velocity than the speed of individual particles~\cite{Kessler2004}. The emergence of such chaotic flow patterns and active turbulence has been observed both experimentally and numerically in a wide variety of active systems~\cite{wensink2012meso,ourSM2015,Urzay17,martinez2019selection,Duclos1120,lin2021energetics}.

Many efforts have been made to explore the analogy between the active turbulence characteristics and classical turbulence in conventional fluids (see~\cite{Alert2020} for a recent review). 
In particular, since a wide range of active systems, such as bacteria and subcellular filaments, are composed of elongated particles, several theoretical and numerical efforts have been directed at studying the active nematics turbulence. Here, in addition to the velocity field the dynamics of orientational order of constituent particles and its coupling to the velocity need to be accounted for and often lead to a chaotic flow state that is interleaved with the chaotic motion of topological defects - singular points in the orientation field~\cite{Doost18}.
For active nematics in a highly viscous regime (low Reynolds numbers), analytical and numerical results of Giomi~\cite{Giomi2015} showed active turbulence is a multiscale phenomenon, identifying a characteristic active length scale and revealing an exponential distribution of vortex areas over a range of scales. Both the existence of active length scale and the exponential distributions of vortex areas were confirmed in subsequent experiments on microtubule-kinesin motor assemblies~\cite{guillamat2017taming,martinez2019selection} and in epithelial cell monolayers~\cite{blanch2018turbulent}. Furthermore, based on the vortex areas scaling and the results of numerical simulations, Giomi proposed a mean field theory describing spectral features of active turbulence, and suggested $E_k \sim k^{-4}$ scaling of the kinetic energy $E_k$ with the wavenumber $k$, without being affected by viscosity and activity~\cite{Giomi2015}.
%
\begin{figure}[bt!]
\centering
\includegraphics[width=\linewidth]{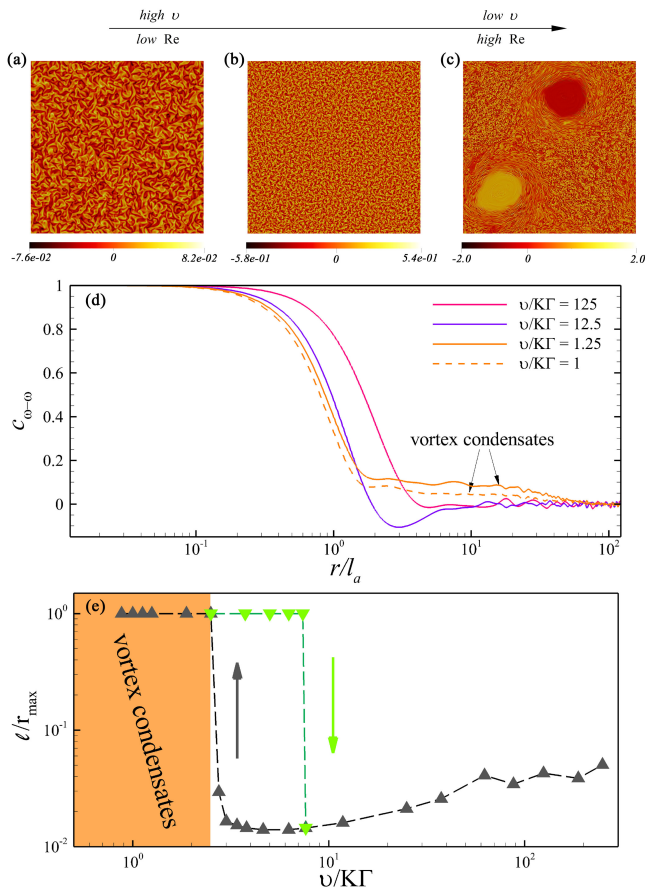}
\caption{{\bf Active turbulence and vortex-condensates}. Snapshots of the flow vortices for incrementally decreasing viscosities: vorticity contours for (a) $\nu/K\Gamma = 125$ with $\rnl\sim0.03$ (equivalent to $Re_\text{T} = \cm{0.0002}$ using the Taylor microscale and \cm{$Re_\text{I} =  \cm{0.005}$ using the integral scale} to define the Reynolds number), (b) $\nu/K\Gamma = 12.5$ with $\rnl\sim0.7$ (equivalent to $Re_\text{T} =  \cm{0.0023}$ using the Taylor microscale and \cm{$Re_\text{I} =  \cm{0.03}$ using the integral scale}), and (c) $\nu/K\Gamma = 1.25$ with $\rnl\sim250$ (equivalent to $Re_\text{T} =\cm{8.2227}$ using the Taylor microscale and \cm{$Re_\text{I} =  \cm{191}$ using the integral scale}). (d) Effect of viscosity on vorticity-vorticity correlations $C_{\omega-\omega}(r)$. The distance $r$ is normalized by the active length scale $l_a = \sqrt{K/\zeta}$. (e) Characteristic vorticity length scale as a function of viscosity. Upon decreasing viscosity, after an initial decrease in the size of vortices, condensates spanning the entire system are formed. The length scale ($\ell$) equals to the length $r$ at which $C_{\omega-\omega}(r)=0$. Green data points represent an incremental increase in viscosity and show the presence of a hysteresis loop, indicating a discontinuous transition to the vortex-condensate state.}
\label{fig:vorticity}
\end{figure}

As with the Giomi's theoretical and numerical analyses, the majority of the previous studies have considered the micro-scale active turbulence, e.g. in cellular monolayers, bacterial suspensions and subcellular filaments-motor protein mixtures, where the Reynolds number is negligible ($\rnl \approx 0$), and viscous dissipation completely dominates over any inertial effects~\cite{alert2021active}. However, in many other realizations of active matter, for example swimming organisms in environmental flows~\cite{dabiri18} and artificial active spinner suspensions~\cite{Kokot12870}, the inertial effect becomes significant, and the Reynolds numbers are non-zero~\cite{klotsa2019above}. 

\begin{table}[h]
		\centering
		\caption{Values of Reynolds number based on different length scales.}\label{tab:re}
		\begin{tabular}{ c c c c } \toprule
			{$\nu/K\Gamma$} & {Active} & {Taylor} & {integral}\\
			{} & {length scale} & {length scale} & {length scale}\\
			{} & {($\rnl$)} & {($\rnl_\text{T}$)} & {($\rnl_\text{I}$)}
			\\ \midrule
			125 & 0.03 & 0.0002 & 0.005\\
			12.5 & 0.7 & 0.0023 & 0.03\\
			1.25 & 250 & 8.2227 & 191\\ \bottomrule
		\end{tabular}\label{tab:Re}
\end{table}
 
Recent studies have begun to reveal interesting impacts of inertia on self-propulsion of active particles and inertial effects on active turbulence~\cite{wang2012inertial,hamel2011transitions,khair2014expansions,scholz2018inertial,lowen2020inertial,linkmann2019phase,koch21,chatterjee2021inertia}. It is shown that increasing the inertia of active particles can result in a transition from active turbulence to flocking in polar active matter~\cite{chatterjee2021inertia}. Moreover, using a one-fluid model of an active matter with hyper viscosity, or a piece-wise constant viscosity, it was found that above a certain Reynolds number active matter can manifest vortex-condensate formation~\cite{linkmann2019phase,condensationactive1} in analogy with the condensates in classical driven 2D turbulence, where inverse energy cascade results in the accumulation of energy at larger scales and condensate formation~\cite{review2Dtrubulence}. More recently, it was shown how the interplay of advective inertia and friction can affect transitions between active turbulence, inertial regime, and tamed inertial active turbulence in active nematic systems~\cite{koch21}. While these studies have provided important insights into the flow features of dense active matter in the presence of inertia, less is known about how inertial effects combined with activity impact the orientational organization of active elongated particles. In particular, singularities in the orientation field, known as topological defects, are increasingly emerging as important centers of self-organization in biological systems~\cite{maroudas-sacks21}, with potential biological functionalities~\cite{saw17,kawaguchi17,meacock21,doostmohammadi2021physics} and how their dynamics are affected by inertial effects is not yet explored.

Here, we report on the numerical investigation of the flow and nematic features of inertial active matter. In order to investigate the fundamental impact of the fluid inertia on the active flow behavior, a continuum model of active nematics is employed. We start by showing the emergence of vortex condensates and then examine its impact on the orientational order and defect density. We then show that not only the defect density, but also the flow around defects get altered within the condensate state and finally show how these combined changes in flow, director, and defect patterns affect energetic features of the active turbulence.
%
\section{\label{sec:method}Methods}
\subsection{\label{sec:gov}Governing equations}
In this study, the two-dimensional continuum model for active nematics is used to model the system's dynamics \cite{Ramaswamy10, Marchetti13,Prost15,Doost18,roozbeh2021}. In this model, the nematic orientations $\hat{n}$, as well as the magnitude of the nematic order $q$, are described with the nematic tensor $\mathbf{Q}$ defined as $\mathbf{Q} = 2q (\hat{n}\hat{n} - \mathbf{I}/2)$. The tensorial definition ensures the nematic (apolar) symmetry of the orientation field ($\hat{n}=-\hat{n}$). The dynamics of $\mathbf{Q}$ is described by Beris- Edwards equation as~\cite{BerisBook}:
\begin{equation}
\frac{\partial \mathbf{Q}}{\partial t}+(\vec{u}\cdot\grad)\mathbf{Q} - \mathbf{S} = \Gamma \mathbf{H},
\label{eqn:lc}
\end{equation}
where $\vec{u}$ is the velocity, and $\mathbf{S}$ is the co-rotation term defined as $\mathbf{S}=\lambda \mathbf{E}-(\mathbf{\Omega}\cdot\mathbf{Q}-\mathbf{Q}\cdot\mathbf{\Omega})$ with ($\mathbf{\Omega}=\frac{1}{2}\crsc{(\grad \vec{u})^\dag-\grad \vec{u}}$) denoting the vorticity tensor and ($\mathbf{E}=\frac{1}{2}\crsc{\grad \vec{u} + (\grad \vec{u})^\dag}$) the rate of strain tensor. Physically, $\mathbf{S}$ describes the response of the orientation field to velocity gradients and the tumbling parameter $\lambda$ determines the degree of this coupling response to the rotational and extensional parts of the flow gradient. The rhs term in Eq.~\eqref{eqn:lc} describes the relaxation of the orientational order to the minimum of the free energy $\mathcal{F}$ through the molecular field $\mathbf{H} = -\frac{\delta \mathcal{F}}{\delta \mathbf{Q}} + \frac{\mathbf{I}}{2} {\rm Tr} \left(\frac{\delta \mathcal{F}}{\delta \mathbf{Q}}\right)$ and is controlled by the rotational diffusivity $\Gamma$. Deformations in the orientation field take place at the cost of a free energy $\mathcal{F}=\mathcal{F}_e+\mathcal{F}_b$ which includes both an elastic free energy ($\mathcal{F}_e=\frac{1}{2}K(\grad\mathbf{Q})^{2}$), penalising gradients in the orientation and a bulk free energy ($\mathcal{F}_b=\frac{A}{2}(1-\frac{1}{2}\text{Tr}[\mathbf{Q}]^2)^{2}$), ensuring a stable nematic ordering at the thermodynamic equilibrium~\cite{santhosh2020activity}. The elastic free energy is approximated by a single elastic constant $K$, and the strength of the bulk free energy is controlled by the coefficient $A$. 

To couple the evolution of the orientation field to the dynamics of the velocity field, generalised incompressible Navier-Stokes equations are considered:

\begin{equation}
\grad \cdot \vec{u} = 0,
\label{eqn:cont}
\end{equation}
\begin{equation}
\rho (\frac{\partial \vec{u}}{\partial t}+(\vec{u}\cdot\grad)\vec{u}) = \grad \cdot\mathbf{\Pi},
\label{eqn:moment}
\end{equation}
which describe the continuity and conservation of linear momentum, respectively.
Here, $\rho$ denotes the density and $\mathbf{\Pi}$ is the stress tensor which includes the viscous term $\mathbf{\Pi}_\text{viscous}=2 \rho\nu \mathbf{E}$ where $\nu$ is the kinematic viscosity, the pressure term $p$,
and active stress $\mathbf{\Pi}_\text{active} =  -\zeta\mathbf{Q}$. Parameter $\zeta$ denotes the activity coefficient and its magnitude controls the strength of the active stress, while its sign determines whether the active particles are extensile ($\zeta>0$) or contractile ($\zeta<0$)~\cite{simha2002hydrodynamic}. 
\cm{Previous simulations of active turbulence at low Reynolds number found that the strength of flows generated by passive stresses could be one to two order of magnitude smaller than activity-induced flows~\cite{Thampi_2015,doostmohammadi2016stabilization}. In this work we neglect the passive elastic stresses in order to establish the defining role of active stresses on the dynamics of an inertial nematic system. Recent studies have considered the potential impact of elastic stresses on active turbulence at low Reynolds numbers~\cite{blanch-mercader2017,carenza2020cascade} and further work is needed to establish their relative importance compared to active and viscous stresses.}

\subsection{\label{sec:sph}Numerical Method}
To solve the coupled governing equations \eqref{eqn:lc} through \eqref{eqn:moment}, a Finite Volume method is utilized based on the open-source OpenFOAM package~\cite{OpenFOAM}. The simulation domain consists of a 2D square of size 200 $\times$ 200, which is discretized using the Cartesian grid with two different resolutions, $1024\times1024$, and $2048\times2048$, and the time step size is controlled through the CFL condition~\cite{cfl,ferziger2002}. Gauss Linear discretization~\cite{moukalled2016} is used for gradient, divergence and laplacian terms, and the PISO algorithm~\cite{versteeg2007} is utilized for the velocity- pressure coupling. The time marching is performed based on the Euler scheme~\cite{ferziger2002}. Periodic boundary condition is enforced on the boundaries. Unless otherwise stated, the parameters used in simulations are according to Tab.\ref{tab:dim}. Since in this study we are mainly interested in the impact of varying the viscosity and activity of the system, dimensionless viscosity $\nu/K\Gamma$ and dimensionless activity $\zeta/A$, are defined. Moreover, Reynolds number is defined as $\rnl=l_a V_\text{rms}/\nu$, where $l_a=\sqrt{K/\zeta}$ is the active length scale~\cite{Giomi2015} and $V_\text{rms}$ is the emergent root-mean-square velocity of the system that varies for different activity and viscosity values. \cm{In order to compare the values of the Reynolds number with the studies of vortex condensate formation in classic turbulence~\cite{gallet2013two,review2Dtrubulence}, the corresponding values of Reynolds numbers are also reported based on the Taylor microscale and the integral scale~\cite{pope2000turbulent} (Fig.~\ref{fig:vorticity}). The Reynolds number based on the Taylor microscale is calculated using $\rnl_{\text{T}}=\lambda V'_\text{rms}/\nu$, where, $\lambda=V'_\text{rms}\sqrt{15\nu/\epsilon}$ is the Taylor microscale with $\epsilon=2\nu<E_{ij}E_{ij}>$ denoting the kinetic energy dissipation rate and $E_{ij}$ represents the strain rate tensor. $V'_\text{rms}$ is the rms of the fluctuating component of the velocity. The Reynolds number based on the integral scale ($\rnl_{\text{I}}$) is calculated using $\rnl_{\text{I}}=L_{\text{I}} V_{\text{I}}/\nu$, where the integral length scale $L_{\text{I}}$ and integral velocity scale $V_{\text{I}}$ are defined as $L_{\text{I}}  = \left. \int_{0}^{\infty} k^{-1} E_k\,dk \middle/ \int_{0}^{\infty} E_k\,dk \right.$, and $V_{\text{I}} = \sqrt{2\kappa}$, respectively~\cite{Urzay17,pope2000turbulent}. Here, $E_k=\tfrac{1}{2}\langle \hat{u}_i(k) \hat{u}_i(k) \rangle$ is the kinetic energy spectrum, $k$ is the wave number, and $\kappa=\langle u_{ij}u_{ij}/2 \rangle$ is the spatially averaged kinetic energy.}

All simulations are initialized with uniform orientation in the horizontal direction subjected to a slight perturbation. Due to the inherently unstable characteristics of active nematics~\cite{Edwards_2009}, these perturbations in the director field lead to the creation of a hydrodynamic instability~\cite{simha2002hydrodynamic}, followed by the nucleation of topological defects and the emergence of active turbulence~\cite{Giomi13,thampi2014}. All quantitative analyses are performed after the active turbulence state is established and the system reaches statistical steady-state.

		\begin{table}
		\centering
		\caption{Values of model parameters employed in the numerical simulation, unless stated otherwise.}\label{tab:dim}
		\begin{tabular}{ c c } \toprule
			{Parameter} & {value} \\ \midrule
			$\Gamma$   & $0.4(m N^{-1}s^{-1})$ \\
			$\lambda$  & $0.7$ \\
			$A$  & $1(Nm^{-1})$ \\
			$K$  & $0.02(Nm)$ \\ 
			$\zeta$ & $0.03(Nm^{-1})$ \\
		$\rho$ & $1(Ns^2m^{-3})$ \\
		    $\nu$   & $[0.01,0.1,1.0] (m^2s^{-1})$ \\ \bottomrule
		\end{tabular}
	\end{table}

\section{\label{sec:res}Results}
\begin{figure}[hbt!]
\centering
\includegraphics[width=\linewidth]{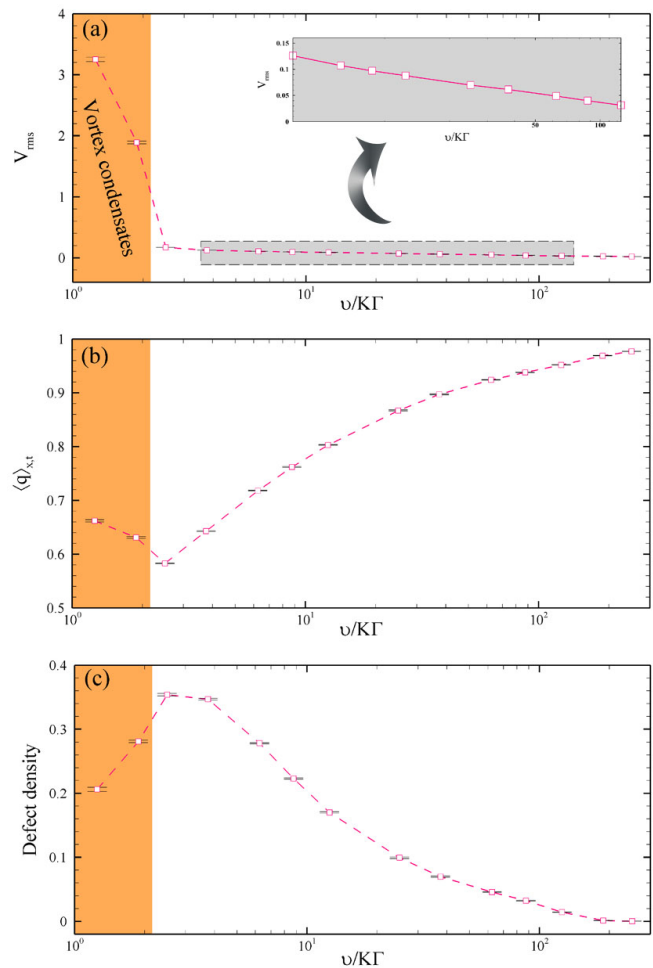}
\caption{{\bf Quantification of viscosity impact on flow and orientation properties of active nematics.} Effect of viscosity on (a) rms-velocity $V_\text{rms}$, (b) magnitude of the nematic order, and (c), defect number, before and after the transition to the vortex-condensate state. Inset in sub-figure (a) shows the semi-log plots of the rms-velocity as a function of the logarithm of viscosity, highlighting its logarithmic decay with viscosity. Similar behavior can be seen for the defect density in (c).} 
\label{fig:correlationANDvrmsdefectqmean}
\end{figure}
We begin by qualitatively assessing the impact of increasing the inertia of the active fluid on the flow patterns by incrementally reducing the kinematic viscosity coefficient.~\figref{fig:vorticity} demonstrates the vorticity contours for three representative viscosities. Upon reducing the viscosity, the size of the vortices becomes smaller, and the strength of vortices is enhanced (Fig.~\ref{fig:vorticity}a,b). Similar to the classical turbulence by reducing the viscosity, and thus increasing Reynolds number, large eddies become unstable and begin to break up, transferring their energy to comparatively smaller eddies~\cite{review2Dtrubulence}. Moreover, in agreement with the reverse cascade dynamics in 2D classical turbulence, a further increase in the Reynolds number results in the accumulation of energy from smaller scales towards the larger scales leading to the emergence of a vortex condensate in the form of two large counter-rotating vortices that span the entire system (Fig.~\ref{fig:vorticity}c). It is important to note, however, that in comparison to the classical inertial turbulence that is driven by external forcing~\cite{review2Dtrubulence}, the transition to vortex-condensate here is driven by active stress generation. The formation of vortex-condensates can best be represented quantitatively through measuring the vorticity-vorticity correlation function $C_{\omega-\omega}(r)=\langle \omega(r).\omega(0) \rangle/\langle\omega(0)^2 \rangle$, where $\omega = \partial_x u_{y} - \partial_y u_{x}$ (Fig.~\ref{fig:vorticity}d). Before and after the emergence of the condensate, different characteristic length scales of decay are exhibited: before the emergence of the condensate, by increasing the Reynolds number, the correlation length decreases, while after the emergence of the vortex condensate, the characteristic length scale is set by the system size that encompasses the two giant vortices (Fig.~\ref{fig:vorticity}d,e).
 
Similar flow patterns of vortex condensate formation were reported earlier based on the one-fluid model of active polar matter with both hyper-viscosity and piece-wise constant viscosity, where it was demonstrated that a discontinuous, subcritical phase transition governs the emergence of the vortex condensate state~\cite{linkmann2019phase,condensationactive1}. By performing the hysteresis analysis, we confirmed that the crossover to the condensate state in inertial active nematics also shows a hysteresis effect, indicating a discontinuous transition to the vortex-condensate state in active nematics (Fig.~\ref{fig:vorticity}e).

It is further shown recently that in active nematics although the energy budget associated with advective inertia could be smaller compared to active and dissipative energies, the effect can accumulate over time leading to large-scale flow patterns~\cite{koch21}.

In addition to the change in the size of the flow patterns, condensate formation is accompanied by a significant increase in the strength of the flow. This can be quantified by measuring the averaged rms-velocity $V_\text{rms}$ of the entire system after reaching a statistical steady-state, which shows up to an order of magnitude enhancement in the velocity upon transition to the vortex-condensate state (Fig.~\ref{fig:correlationANDvrmsdefectqmean}a). Moreover, a closer look at the variation of the velocity beyond the transition point into the vortex-condensate state reveals a logarithmic decay of the rms-velocity with viscosity, as exemplified by the semi-log plot in the inset of Fig.~\ref{fig:correlationANDvrmsdefectqmean}a.
\begin{figure}[hbt!]
\centering
\includegraphics[width=\linewidth]{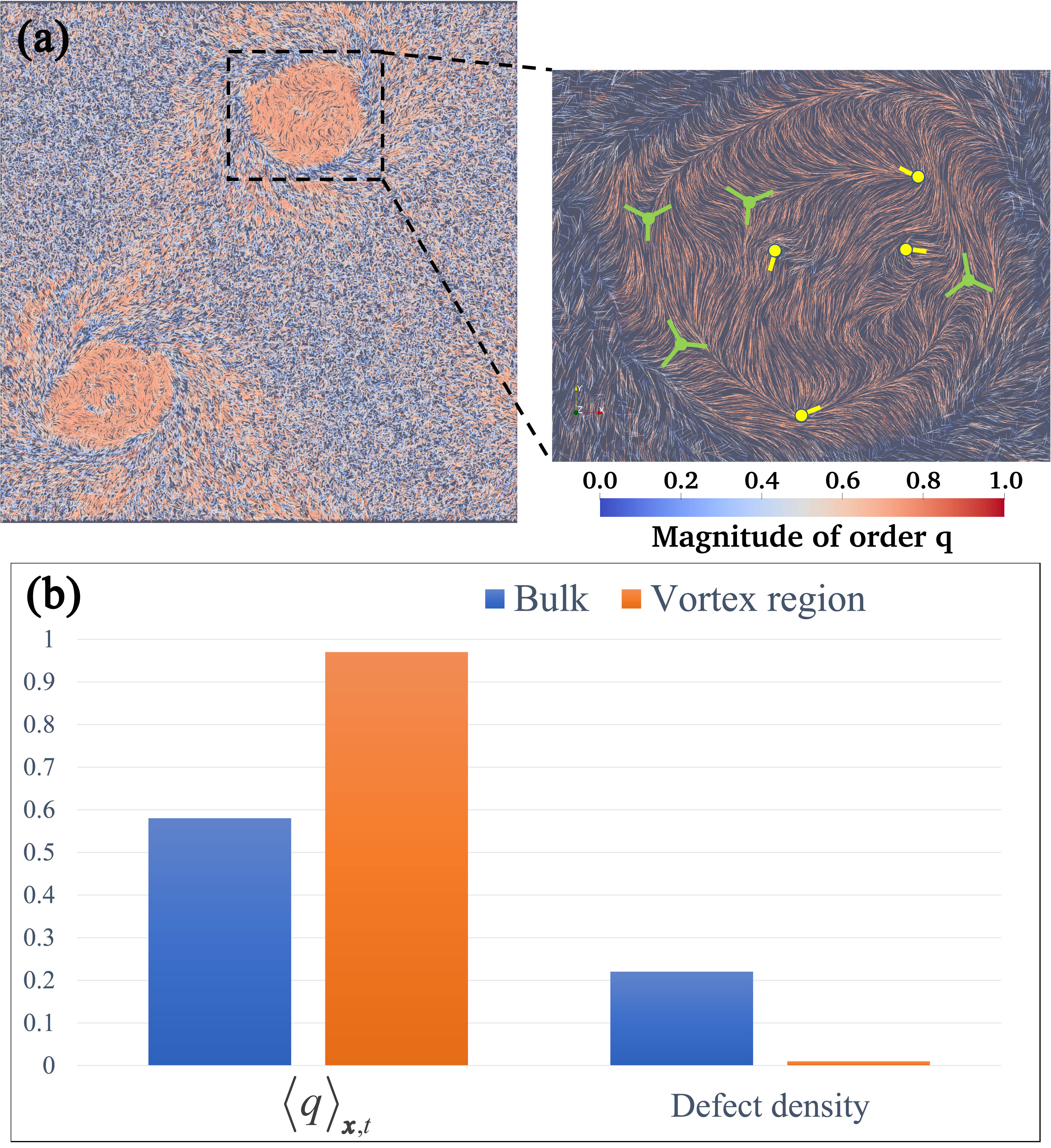}
\caption{{\bf Nematic order condensation}. (a) snapshots of director field and topological defects for vortex condensates case. Colormap indicates the magnitude of the nematic order $q$, and $+1/2$ and $-1/2$ topological defects are marked by yellow comets and green triangles, respectively.(b) Averaged values of the magnitude of order $\langle q \rangle_{{\bf x},t}$, and defect density calculated separately inside giant vortices and the bulk of the system excluding the giant vortices.}
\label{fig:snapshot}
\end{figure}

We next asked what would the consequences of such drastic changes in the strength and patterns of the flow be on the orientation field of the active nematic particles. To test this, the magnitude of the nematic order averaged over time and space $\langle q \rangle_{{\bf x},t}$ was measured for incrementally decreasing values of viscosity (Fig.~\ref{fig:correlationANDvrmsdefectqmean}b). Interestingly, the initial melting of the nematic order before the condensate formation is followed by an increase in orientational ordering within the vortex-condensate state. This is further accompanied by changes in topological defects density within the system, which after an initial increase with decreasing viscosity, begins to fall as the vortex-condensate is established (Fig.~\ref{fig:correlationANDvrmsdefectqmean}c). A closer look at the director field associated with the vortex-condensate state reveals the underlying mechanism for such changes in the nematic order and topological defect density: once the vortex-condensate forms, within the two giant vortices, a nearly perfect nematic order is established that is only disrupted by few topological defects (Fig.~\ref{fig:snapshot}a), while the bulk of the system is characterized by disordered domain laden with a high density of the topological defects. As a result of this {\it order condensation} within the giant vortices, the magnitude of the order increases within the condensate phase, which is accompanied by a drop in the total defect density. 

To explain the reason for the reduction/increase in defects population/orientational order at the condensate state, we quantified the nematic order within and outside of the vortex condensate region, showing clearly that the emergence of giant vortices is accompanied by the enhancement of the order and thus fewer defects within the condensates (Fig.~\ref{fig:snapshot}b).
The mechanism for this can be demonstrated in a simplified form by approximating a giant vortex as an ideal Rankin\cm{e} vortex with the velocity profile in the polar coordinate $(u_{r}=0, u_{\theta}, u_z = 0)$, with:
\begin{equation}
u_{\theta}=\frac{\Lambda}{2\pi}
    \begin{cases}
    r/a^2,& r \le a\\
    1/r, & r > a
    \end{cases}
\end{equation}
where $\Lambda$ is the strength of the circulation of Rankin\cm{e} vortex and $a$ is the vortex core size. This results in finite vorticity $\omega_z = \Lambda/(\pi a^2)$ \cm{and solid body rotation} in the core region. \cm{An approximately constant vorticity across the giant vortex region is evident from the snapshots of the vortex-condensate (Fig.~\ref{fig:vorticity}c). Previous works have established that `walls' of large nematic distortion and topological defects are typically formed in the regions between separate vortices~\cite{Giomi2015,thampi2016active}. Therefore, here, as a result of the constant vorticity and an approximate solid body rotation within the giant vortex there is only a weak destabilizing effect from variations in vorticity} to frustrate the nematic order and as such higher order is expected within the vortex condensates.

\begin{figure}[hbt!]
\centering
\includegraphics[width=\linewidth]{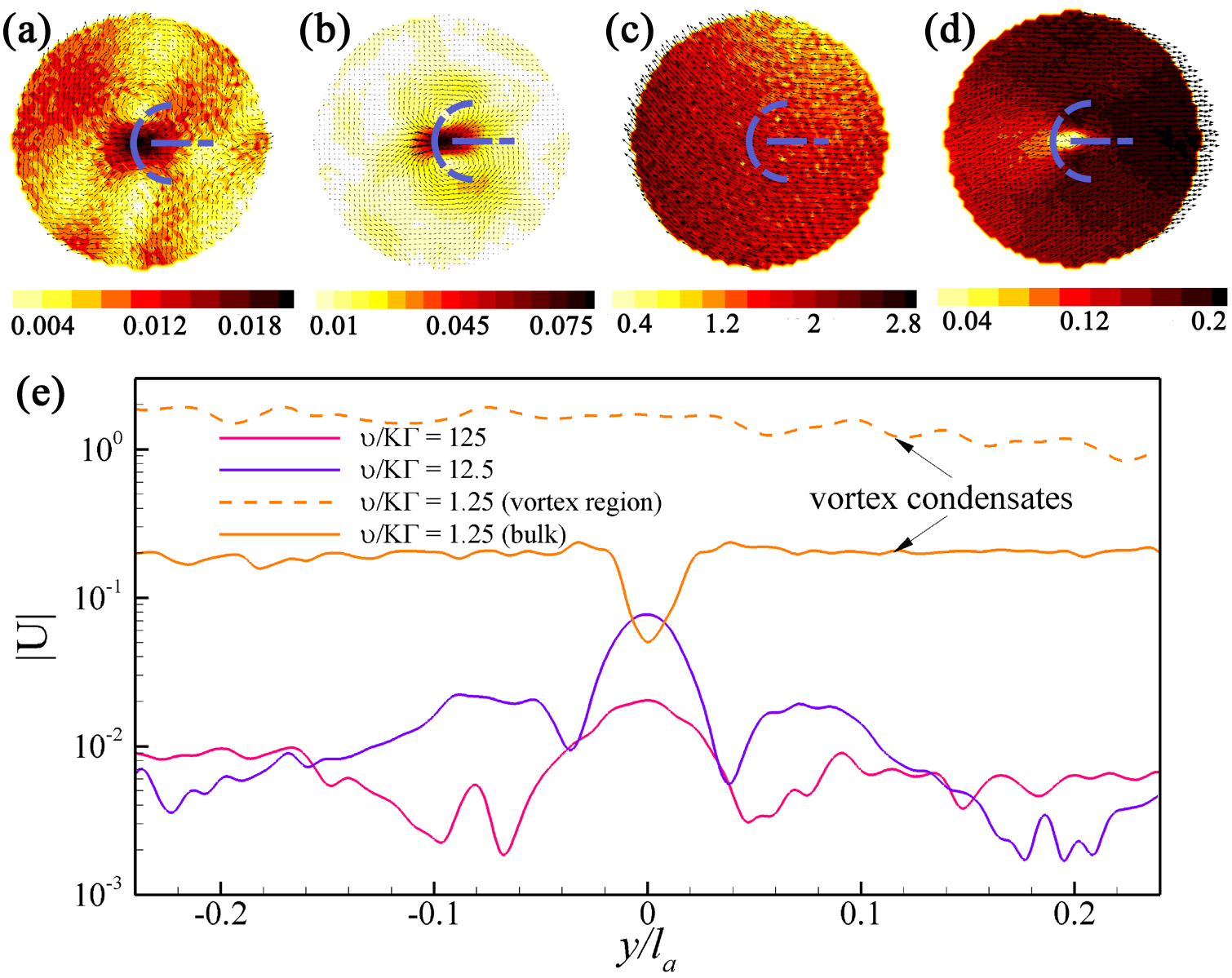}
\caption{{\bf Viscosity impact on the flow field of topological defects.} Velocity field and contours of velocity magnitude for the average defect flow at (a) $\nu/K\Gamma = 125$ and (b) $\nu/K\Gamma = 12.5$. (c) and (d) show the average defect flow for the vortex-condensate state at $\nu/K\Gamma = 1.25$, calculated separately for (c) defects inside giant vortices and (d) the bulk of the system excluding the giant vortices. In (a)-(d) blue dashed line schematically show the alignment of the $+1/2$ defect with respect to the averaged flows. (e) Velocity profile around defects showing the magnitude of velocity along a vertical axis passing through the center of defects in (a), (b), (c) and (d).}
\label{fig:Avdefectflow}
\end{figure}
As such, the orientational order and topological defects display different features within the giant vortices and the bulk of the system in the condensate phase. To gain further insights into the potential impact of varying viscosity and also the distinction between the bulk and the vortex region, we measured the average flow around topological defects. It is well-established that in active turbulence with negligible inertia, comet-shaped $+1/2$ defects show propulsive motion within the system, as evident from the average flow at the highest value of the viscosity, which is consistent with analytical predictions using Green's function and experimental measurements in dense cellular systems~\cite{Giomi2015,saw17,meacock21} (Fig.~\ref{fig:Avdefectflow}a). Upon decreasing the viscosity the flow field around the defects keeps its shape, however, the size of the flow vortices around the defect is reduced and the flow strength at the defect core is enhanced (Fig.~\ref{fig:Avdefectflow}b). This means that at lower viscosities the speed of the propulsion of $+1/2$ defects increases, while their higher density leads to a more effective screening of their associated flows due to the stronger interactions with other defects. 

Further decrease in viscosity and the emergence of the vortex-condensate, however, completely alters flow features of the $+1/2$ topological defects: within the giant vortices the defects move along their comet head and rotate around the vortex center as characterized by the tilted averaged velocity field (Fig.~\ref{fig:Avdefectflow}c). Remarkably, within the bulk of the condensate phase, the magnitude of the velocity also drops at the defect core and the average flow of the $+1/2$ defects points along the defect tail indicating that the defects align anti-parallel to the flow direction (Fig.~\ref{fig:Avdefectflow}d). This is because within the bulk, the strong shear flow between the two giant vortices aligns $+1/2$ defects anti-parallel to the strong flow and the defects are advected by the strong flow between the two vortices. As such within the condensate phase, the propulsive nature of the $+1/2$ defects is suppressed by the flow field established through vortex-condensate and defects have negligible impact on the flow field, contrary to the active turbulence state where the propulsive nature of the $+1/2$ is an indispensable determinant of the flow structure within the system~\cite{Thampi_2015,Giomi2015}. The flow characteristics around defects can be further quantified by calculating the decay of the velocity magnitude away from the defect core, which clearly demonstrates the alterations to the flow scale with decreasing viscosity and with the emergence of the vortex-condensate phase (Fig.~\ref{fig:Avdefectflow}e).

\begin{figure}[h!]
\centering
\includegraphics[width=\linewidth]{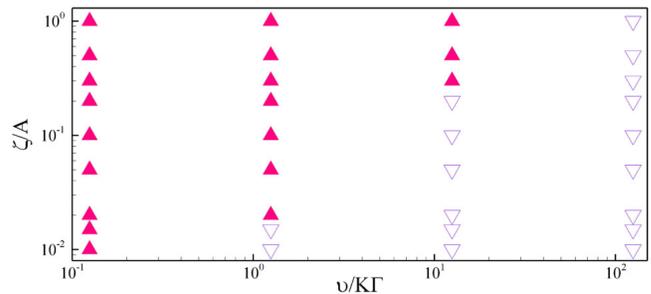}
\caption{{\bf Stability diagram for the vortex-condensate formation in active nematics.} Effect of viscosity and activity on the vortex condensation formation are shown. Filled points represent the vortex condense state.}
\label{fig:activityviscosity}
\end{figure}

It is noteworthy that the emergence of the vortex-condensate depends not only on viscosity, but also on the activity of the particles. To show this, \figref{fig:activityviscosity} illustrates the stability diagram of the vortex-condensate formation in the viscosity-activity phase space. As evident from the figure, even at moderately high viscosities, vortex-condensate can form for strong enough activities. Only at significantly high viscosity, where the inertial effects are completely suppressed by the viscous dissipation and the convective inter-scale transfer is insignificant, no vortex-condensate is observed.
Moreover, the emergence of the vortex-condensate is only observed for extensile (pusher) particles and even for high values of contractile activity and strong inertia the active turbulence is established and we could not find any condensate state. We conjecture that the observed difference between the extensile and contractile systems can be associated with the difference in the response of the collection of extensile and contractile active particles to the flow gradients: extensile particles collectively align to the shear flows established by their self-generated active stresses, which leads to the local ordering of extensile active systems, as shown previously~\cite{Thampi_2015,santhosh2020activity}. On the other hand, contractile activity destroys such an ordering. This effect is best evident from previous studies on active nematics that show combined effects of activity and flow alignment lead to renormalization of the molecular field, as shown in the context of intrinsic free energy of active nematics~\cite{Thampi_2015}: for non-zero and positive values of the flow-aligning parameter $\lambda$, contractile $\zeta < 0$ and extensile activities $\zeta > 0$ have the opposite impact on the effective free energy of the system. As such, contractile activity increases the energetic cost of the breakdown of nematic order, while extensile activity enhances it.

\begin{figure}[h!]
\centering
\includegraphics[width=\linewidth]{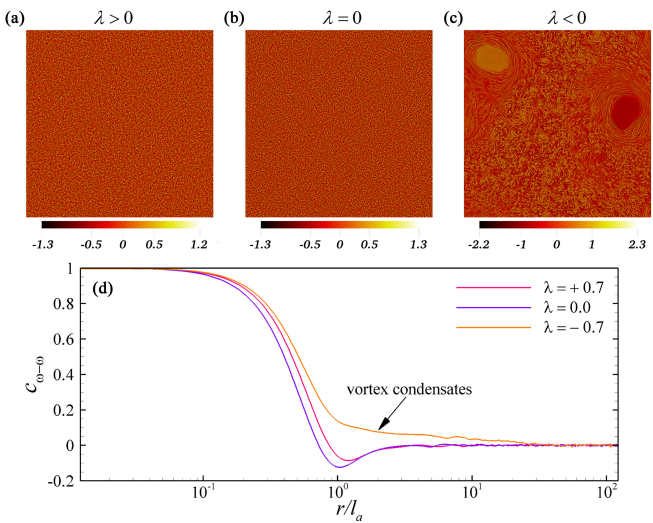}
\caption{{\bf Active turbulence and vortex-condensates for contractile activity ($\zeta/A = -0.03 $) at $\nu/K\Gamma = 1.25$}. Snapshots of the flow vortices for positive, zero, and negative tumbling parameter: vorticity contours for (a) $\lambda = +0.7$, (b) $\lambda = 0$, and (c) $\lambda = -0.7$. (d) Effect of tumbling parameter on vorticity-vorticity correlations.}
\label{fig:contr}
\end{figure}

To test this conjecture directly in our simulations, we explored cases with contractile activity ($\zeta < 0 $) and negative values of the flow-aligning parameter ($\lambda < 0$). The results confirm that it is possible to obtain a vortex-condensate state for contractile activities when the flow-aligning parameter is negative and emphasize that the sign of the product $\zeta \lambda$ is the determining factor (Fig.~\ref{fig:contr}). Furthermore, for $\lambda = 0$ there was no vortex-condensate state for neither contractile nor extensile activity.

Having established the impact of viscosity reduction on the flow and director field of active nematics, we next turn to the energetic features of the flow. This is best represented by the kinetic energy spectrum $E_k=\tfrac{1}{2}\langle \hat{u}_i(k) \hat{u}_i(k) \rangle$, which measures the kinetic energy associated with differing scales characterized by the wavenumber $k$. A numerical study of the simplified active nematics, which neglects order variation and thus topological defects, suggested a universal scaling of the kinetic energy, $E_k \sim k^{-1}$ at small wavenumbers~\cite{Alert2020}. While such a $k^{-1}$ scaling is recently observed in a numerical study of active {\it polar} fluid in certain parameter regimes~\cite{chatterjee2021inertia}, numerical simulations of the full active nematics did not find such universal behavior~\cite{Urzay17,ravnik2019,amiri2021}. Most recently, a combined theoretical and experimental study showed different scaling regimes depending on the external or internal dissipation mechanisms for microtubule-kinesin motor mixtures at oil-water interface, which represent a realization of two-dimensional active nematic material~\cite{martinez2021scaling}.

\begin{figure}[t!]
\centering
\includegraphics[width=\linewidth]{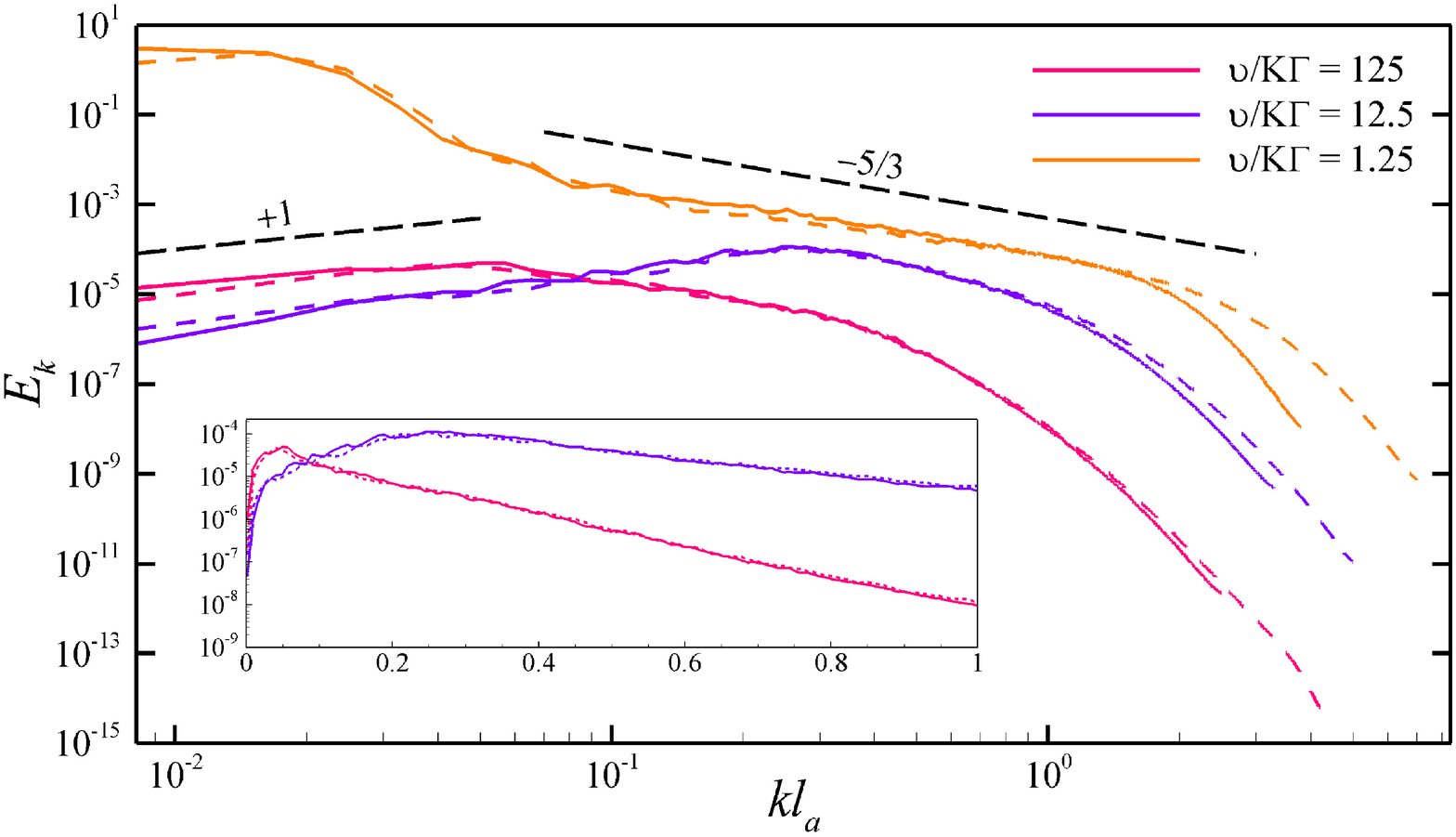}
\caption{{\bf Kinetic-energy spectra.} The wavenumber is non-dimensionalized by the active length scale $l_a = \sqrt{K/\zeta}$. Solid- and dashed- lines, respectively represent $1024 \times 1024$ and $2048 \times 2048$ grid resolutions. While within the vortex-condensate state a power-law decay is observed, lower viscosities manifest exponential decay with the wavenumber (semi-log plots in the inset).}
\label{fig:energy}
\end{figure}

Let us first consider the vortex-condensate case that appears at low viscosities, corresponding to high Reynolds numbers. In agreement with classical 2D turbulence within the inertial range the power spectrum shows Kolmogorov scaling with a power-law decay with the exponent $-5/3$~\cite{review2Dtrubulence} (\figref{fig:energy}; {\it orange line}). This indicates that at low viscosities, the vortex-condensate generated by local energy injection in active nematics shares similar scaling behavior as driven inertial turbulence. Incremental reduction of inertial effects by successively increasing the viscosity, however, completely alters the scaling behavior of the active nematic turbulence. At small wavenumbers, corresponding to large scales, the kinetic energy shows again a power-law behavior with a non-universal viscosity-dependent exponent (\figref{fig:energy}; {\it purple and magenta lines}). The transfer of energy to smaller scales (larger wavenumbers), however, does not follow a universal power-law decay, as has been suggested by analytical and numerical studies of active nematics that neglect topological defects~\cite{Giomi2015,Alert2020}. Instead, representing the energy spectrum on a semi-log plot reveals a viscosity-dependent exponential decay of the energy with wavenumber, that is due to the dominating effect of viscous dissipation with reducing the inertia. This is important, because existence of universal scaling laws for active turbulence, that are independent of activity and viscosity, have been suggested based on power-law decay of energy spectrum with the wavenumber as $\sim k^{-4}$~\cite{Alert2020}. The large-scale numerical simulations conducted here that account for topological defects in the system do not show any indication of such universal scaling laws and rather suggest a dissipation dominant exponential regime at low Reynolds numbers, calling for further studies in this direction.
\section{\label{sec:conc}Conclusions}
The effect of fluid inertia on the self-organization of active nematic systems has been largely overlooked. Here, we provide finely-resolved simulations of the active nematohydrodynamic equations in the presence of progressively increasing inertial effects. By incrementally reducing the fluid viscosity we quantify the variations of emergent system properties including velocity, orientational order, and density of topological defects. At sufficiently small values of viscosity the system manifests reverse energy cascade and formation of vortex-condensate, as well as the Kolmogorov scaling in the decay of the kinetic energy, similar to classical driven turbulence, albeit at significantly lower Reynolds numbers, \cm{$Re_{\text{I}}\sim O(10^2)$ (defined based on the integral scale)}, and due to the local energy injection by activity, rather than any external forcing. We provide a phase diagram of the vortex-condensate formation in the activity-viscosity phase space, emphasizing the synergistic impacts of active stresses and fluid inertia. Our results further show that vortex-condensate formation is accompanied by significant modifications of the orientational features of the system, leading to order condensation and a drop in the defect density. Analyzing the defect within the condensate state, further demonstrates a significant deviation in the flow field around the topological defects.Importantly, we show that the condensate-formation in inertial active nematics depends strongly on the flow-aligning behavior of the active particles and the extensile or contractile stresses that they generate.

Our results further reveal that in the non-condensate regime, the averaged velocity of the entire system shows logarithmic decay with the fluid viscosity. Additionally, within this regime, we show that the kinetic energy spectrum lacks any universal scaling. Indeed, the results even at a highly viscous regime indicate an exponential - rather than algebraic - decay of the energy spectrum with the wavenumber.

The results presented in this study demonstrate the important interplay between the active stress generation and fluid inertia in active fluids. While the bulk of the studies of active matter in general, and active nematics in particular, have focused on the highly viscous regime, applicable to bacterial suspension and subcellular filaments, where inertial effects are absent, fluid inertia is expected to play a role in collective organization of larger swimming organisms~\cite{klotsa2019above} such as marine zooplanktons Copepods and brine shrimp ({\it Artemia salina}) that are commonly encountered in environmental fluids~\cite{van2003escape,katija2009viscosity,dabiri18}. Within such inertial active fluids, the observed changes in averaged velocity, orientational order, topological defect density, and energy spectrum could result in significant alterations in collective behavior and foraging traits of the self-propelled organisms~\cite{kiorboe2009mechanistic,doostmohammadi2012low}. Moreover, it would be interesting to probe the impact of inertia in synthetic realizations of active nematics, where the viscosity of the medium can be directly tuned to explore the crosstalk between activity-induced and inertial effects on the turbulent pattern formation of active particles.

\section*{Acknowledgements}
A. D. acknowledges support from the Novo Nordisk Foundation (grant No. NNF18SA0035142), Villum Fonden (Grant no. 29476), and funding from the European Union’s Horizon 2020 research and innovation program under the Marie Sklodowska-Curie grant agreement No. 847523 (INTERACTIONS).

\bibliography{apssamp}

\end{document}